\documentstyle [12pt,twoside]{article} 
  
\begin{document}
\begin{flushright}
UBCTP--96--001
\end{flushright}
\centerline{\large{\bf A Macroscopic Gravity Wave Effect}}
\vspace*{1.cm}
\centerline{\bf Redouane Fakir}
\vspace*{0.5cm}
\centerline{\em Cosmology Group, Department of Physics}
\centerline{\em University of British Columbia}
\centerline{\em 6224 Agriculture Road, Vancouver, B.C. V6T 1Z1, Canada}
\centerline{\em fakir@physics.ubc.ca}
\vspace*{1.5cm}
\centerline{\bf Abstract} 
\vspace*{1.cm}

Gravitational waves, although generally associated with
extremely microscopic effects, can displace by hundreds of kilometers
the pulsar interstellar scintillation patterns that 
bathe the Earth. The combination 
of the pulsar and the interstellar medium acts as a
kiloparsec-long, nature-provided gravity wave amplifier.  
We show how an effective scheme for the 
detection of periodic gravity waves can be constructed based 
on this effect. This approach to detection does not require the development
of new, ad hoc technology, but the optimization of existing 
observational techniques in a few different fields of astronomy.
Part of the scheme is a new, purely numerical detection
technique that can also be used in the data processing of 
other projects of periodic gravity wave detection.

\clearpage
It was recently realized that gravity waves can interfere
with interstellar scintillation in a potentially observable way [1]:
Pulsar radio waves are scattered by electron density
inhomogeneities in the interstellar medium. This results in a pattern
of intensity maxima and minima of typical size ${\bar S}$ in the
vicinity of the Earth. As the latter moves through the pattern with
a relative velocity $V_{r}$, the average pulse intensity appears
to fluctuate on a timescale ${\bar t_{s}} = {\bar S}/V_{r}$. This
is interstellar scintillation. (For an excellent review and references, 
 see [2,3].) When a foreground gravity wave source
such as a binary star lies close  enough to the pulsar line-of-sight,
the steering effect of the gravity waves [4,5] moves the scintillation pattern
quasi-rigidly and periodically by a distance $D=\alpha_{gw} L$,
where $\alpha_{gw}$ is the deflection angle due to the close
encounter of the radio waves  with the gravity wave source, 
and $L$ is the distance from that source to the Earth.

This amounts to a natural amplification of the gravity wave deflection effect by
the gigantic lever arm $L$. A preliminary search by C.F. Quist [6] has  
identified several actual cases of pulsar-binary star alignment for which the displacement $D$ could reach hundreds of kilometers (examples below.)
Hence the perhaps unexpected possibility that gravity waves
could be responsible for natural periodic phenomena that are macroscopic.

Several approaches  to the detection of this effect have been 
envisaged, two of which are discussed here (schemes 1 and 2), 
the second of the two
being experimentally the most advantageous. 
In both schemes, part of the detection is done purely numerically
in a manner that is akin to stochastic resonance. 
Under certain conditions,
the noise in a {\em nonlinear dynamic} system, instead of being a hindrance
to signal detection,
can be used to reveal the presence of a weak (weaker than the noise)
periodic signal. This is stochastic resonance [7].
 Paradoxically, the signal in some such systems would not
be detectable at all in the absence of noise. In our case, the raw data
is the noisy pulsar intensity time series $I(t)$, which is a random distribution
reflecting the passage of the Earth through the irregular scintillation
pattern (fig.1). In the following, we turn the stochastic resonance argument
around to obtain a method for detecting weak periodic gravity wave 
signals. This method is independent of (and often more efficient than) the 
usual period folding of data strings. The method can also be adapted
to almost any other scheme of periodic gravity wave detection.

However, before stochastic resonance ideas can be used to detect the
gravity wave effect above, two problems need to be solved:
First, gravity waves modulate the {\em phase} of the intensity fluctuations
(${\tilde I}(t) = I(t+\delta t)$, where ${\tilde I}(t)$
 is the intensity in the presence
of gravity waves) whereas a priori stochastic resonance ideas
only apply to {\em amplitude} modulations (problem 1.)
Second,
stochastic resonance is usually shown to arise from a nonlinearity
in the system's dynamics, but no such dynamics is at play in 
our case (problem 2.)
We shall study two
different solutions to problem 1, each leading to a different
detection scheme. Problem 2 is then addressed separately within
each scheme.

{\em Scheme 1}.- Our first approach to solving problem 1 draws on 
some old and well 
established methods developed in connection with threshold systems
such as electric relays [8,9,10]. We set a pulsar intensity threshold 
$I_{0}$ and decide that a peak of intensity is defined by 
${\tilde I}(t) > I_{0}$ (fig.1). This introduces a new stochastic variable:
the peak duration $\tau (t)$. It can be shown (e.g. [10]) that this new 
variable has well defined statistical properties that are derivable
from those of the original variable ${\tilde I}(t)$, provided $I_{0}$ 
is chosen to be sufficiently large. Since, as seen earlier, the
gravity wave induced displacement of the pattern modulates the time
spent by the observer inside any particular intensity feature, it directly
modulates the {\em amplitude} of $\tau(t)$. 
Thus, the threshold mechanism has turned the 
phase modulations of ${\tilde I}(t)$ into amplitude modulations of $\tau(t)$,
which solves problem 1. 

To build a detection scheme around this particular
solution to problem 1, we still need to address problem 2:
the absence of nonlinear dynamics. The threshold mechanism
again provides an answer. This time it is applied to $\tau(t)$
itself. Since $\tau(t)$ is modulated in {\em amplitude}, one 
can devise a purely numerical procedure (based on a chosen threshold
$\tau_{0}$) that is completely analogous to a physical trigger
mechanism. Trigger mechanisms being obviously highly nonlinear,
this has the potential of solving problem 2.

Let us then arrange that our computer code ``fires'' (i.e. registers
a pulse, see [9,10]) whenever $\tau(t) > \tau_{0}$. The result is yet
another stochastic distribution: a random pulse train $p(t)$. The pulses are
square functions with an arbitrary fixed hight and an 
arbitrary fixed small width (fig.1.) 

The argument now is
that this nonlinear pseudo-dynamic system does exhibit stochastic
resonance. That is, if the irregularities in the raw intensity data
${\tilde I(t)}$ are very weakly but coherently phase modulated by gravity 
waves, this modulation eventually shows up as a periodicity in the pulse
train $p(t)$, which is the output of the double threshold mechanism
just described (applying thresholds $I_{0}$ then $\tau_{0}$; see fig.1.)

That this is indeed the case can be shown analytically by applying
the results in [8,9,10] to the case of a weak cosine signal of
amplitude $s$ and angular frequency $\omega_{gw}$
buried in Gaussian white noise, which is a good approximation
to the physical case at hand [3]. The statistics of the pulse train are
obtained from the correlation function $\psi(t) \equiv <\tau(0)\tau(t)>$.
The rate at which $\tau(t)$ crosses the threshold 
$\tau_{0}$ is [9,10]
\newline\begin{equation}
R(\tau_{0}) = {1\over 2\pi} \left( -{\psi''(0)\over\psi(0)}
e^{-\tau_{0}^{2}/\psi(0)}  
\right)^{1/2} \  \ ,
\end{equation}\newline
where primes indicate time derivatives.
The power spectrum $\cal{P} (\omega)$ of $p(t)$ can be calculated either 
directly or by Fourier transforming the correlation function 
$<p(0)p(t)>$ (i.e. using the
Wiener-Khintchine theorem.)  

The signal-to-noise ratio ($SNR$) can then be obtained 
by deviding the power at the signal frequency $\cal{P}$$(\omega_{gw})$
by the power of the output background noise interpolated at $\omega_{gw}$.
We thus obtain 
\newline\begin{equation}
SNR = {1\over \sqrt{3}} {\Delta t\over t_{c}} {s^{2}\over \sigma^{2}} 
{\tau_{0}^{2}\over \sigma^{2}} 
\exp \left( -{\tau_{0}^{2}\over 2\sigma^{2}} \right) \  \ ,
\end{equation}
where $\sigma$ is the
standard deviation of the random noise, 
$\Delta t$ the total duration of the experiment and
$t_{c}$  the 
average duration of a single measurement.
The ${\sqrt 3}\ t_{c}$ term comes from the Rice
threshold crossing rate for Gaussian white noise [9].
It can easily be verified that the results are 
qualitatively the same for other types of noise,
so long as the correlation time of the noise is much smaller
than the period of the deterministic signal (which is the case
in our observational situation.) 
In the case of
scheme 1, $t_{c}$ is of the order of the 
correlation time of the $I(t)$ sequence.
Eq.(2) was verified numerically (see below.)

Hence, since $\Delta t >> t_{c}$, the mere tuning of the threshold $\tau_{0}$ 
(a trivial and purely numerical procedure)
can make $SNR > 1$ even if the noise is much louder
than the periodic signal ($\sigma >> s$). In stochastic
resonance, it is usually the noise level that is adjusted
in order to achieve detection.
In our case, where the noise level $\sigma$ is a fixed observational 
constraint,  we maximize $SNR$ by adjusting
the {\em threshold} $\tau_{0}$   ($\tau_{0} \approx \sigma$.)
In this way, one  
can a priori detect any periodic signal with an amplitude 
exceeding $s_{min} \approx \sigma\sqrt{t_{c}/\Delta t}$. 
For very weak signals, some fine tuning
of $\tau_{0}$ is necessary when confirming this numerically.

To maximize the action of the interstellar medium, 
which is the master piece of the ``natural'' detector,
one should observe using the lowest possible electromagnetic
frequencies. This means observing at about $50 MHz$,
which corresponds to $t_{c} \sim 1 min$ for several pulsars [2,3].
On the other hand, a realistic value for the total duration
of the experiment is $\Delta t \approx 10^{8}sec$.
These numbers result in $s_{min} \approx \sigma/10^{3}$.
Thus, this mechanism is in principle capable of detecting
signals that are $1000$ times weaker than the noise.

In practice, this sensitivity will surely be reduced by the variance 
of the periodigrams $P(\omega)$ which are actually calculated and which
approximate the power spectrum ${\cal P}(\omega)$. We shall return to this point
further below when we apply similar techniques to the more
sensitive scheme 2. However, two remarks should be made here.
(1) The detection procedure described above is independent of the
gravitational wavelength: it can be applied to detect gravity waves
with periods $T_{gw}$ anywhere between $T_{gw} = t_{c}$ and 
$T_{gw} = \Delta t$. (2) The above procedure
is unrelated to the usually smaller sensitivity increase (about
$\sqrt{\Delta t/T_{gw}}$) that can be achieved by folding the
data over the gravity wave period $T_{gw}$. For binary stars,
$\sqrt{\Delta t/T_{gw}}$ is usually smaller than $10$, while
numerical simulations show that the above detection technique
is sensitive {\em in practice} at the $\sigma/s\approx 100$ level, 
even before any periodigram
variance reduction. For the fastest binaries ($T_{gw}\sim 1 hour$),
the theoretical sensitivity increase   
from period folding  $\sqrt{\Delta t/T_{gw}}$
is still about one order
of magnitude smaller than the theoretical sensitivity increase 
from the above method $\sqrt{\Delta t/t_{c}}$.
 In fact, folding the data first and {\em then} using
the new method represents a double sensitivity enhancing procedure
(operating in the time domain) that is an alternative to  
applying the new method first as was done above, and then performing a periodigram
variance reduction procedure (thus operating in the frequency domain.)

We have learned recently that the strength of this gravity wave numerical detection method is in fact a manifestation of a broader
phenomenon (dubbed non-dynamic stochastic resonance) discovered
lately by F. Moss [11] quite independently from
gravity wave considerations. 
Note however that the signal-to-noise ratio formula eq.(2)
above is somewhat different from its counterpart in [11]: 
(1) it is dimensionless
and (2) it increases with the total observation time $\Delta t$.
Moreover, motivated by specifically gravity wave constraints, our 
numerical code is designed to detect periodic signals that are 
much weaker with respect to the noise, and also, it does not take advantage
of periodigram averaging. (More details below.) 

{\em Scheme 2}.- We now go back to the problem that gravity waves modulate the
phase rather than the amplitude of the pulsar intensity data (problem 1).
The detection scheme above (scheme 1) was based on solving problem 1
by deriving from $I(t)$ the peak duration variable $\tau(t)$ which {\em is}
amplitude modulated. In the following we investigate a different
solution to problem 1, which turns out to generate a more sensitive
detection scheme (scheme 2). One more astronomical fact will now be
exploited, albeit a trivial one: the fact that the ground based
observer has
access to a finite space of the order of the Earth diameter $D_{E}$.

A perhaps remarkable coincidence makes this fact not so benign:
$D_{E}$ is of the same order of magnitude as the typical size of
intensity features in the interstellar scintillation pattern (at the
electromagnetic frequency of $50 MHz$ chosen here.) This makes it fruitful
to register the {\em same} intensity feature more than once, 
as it passes over two or more widely spaced radio telescopes. 
The relative velocity of the scintillation pattern can then
be deduced directly from the cross-correlation of the intensity
records collected at the different stations (see [2] and references therein.)
When gravity waves are at work, the pattern velocity becomes 
${\tilde V}(t) = V(t) + n(t) + v_{gw}(t)$, where $v_{gw}(t)$ is 
the gravity wave contribution and $n(t)$ is the noise from all other
sources. 
Hence, the gravity wave effect generates
here an {\em amplitude} modulation at the outset, 
which solves problem 1. 

The virtual trigger mechanism applied
to $\tau(t)$ in scheme 1 can now be applied to 
$({\tilde V}(t) - V(t))$ (after normalization by subtraction of
the mean), thus solving problem 2.
Here, the pulse width could be made variable and set equal successively
to the time intervals during which $({\tilde V}(t) - V(t))$ is above
the chosen threshold. 
However, numerical simulations show that such a change in the pulse shape does not affect the results qualitatively;
i.e., the key information here is contained in the {\em separation} between 
pulses.

The great advantage of scheme 2 over scheme 1 is that the noise
in scheme 2 (the uncertainty in the normalized $({\tilde V}(t) - V(t))$)
can be substantially reduced by various precision
techniques of scintillation pattern velocity measurements,
whereas the noise in scheme 1 (the variation in the peak duration $\tau$
from one feature to the other) was mostly due to the stochastic
nature of the interstellar scintillation, about which little
can be done.
With the specific aim of gravity wave detection in mind,
the precision in pattern velocity measurements can be substantially
improved over what it has been so far in interstellar scintillation
experiments, which had rather different priorities and objectives.
For example, experiments such as the Penticton-Jodrell Bank investigation
of PSR 0329+54 were often interested in the exceptional cases of pulsars with large proper velocities or
lines-of-sight with  high levels of interstellar medium instability [2,3].
This is just the opposite of the slow and quiet conditions that
would be favorable to the present gravity wave detection schemes. 

Let us recapitulate the steps that are involved in detecting
periodic gravity waves according to scheme 2. Some of these
steps will demand the optimization of existing astronomical
techniques, but none will demand the development of
completely new ad hoc technology.

First, one must investigate the set of known pulsars
(which are rapidly approaching $1000$) for alignment
with foreground binary stars. C.F. Quist has conducted
a very preliminary run of $500$ pulsars against only {\em catalogued}
stars [6]. $106$ objects were found at less than $2'arc$
of pulsar lines-of-sight. About half of these objects are
expected to be binary stars. Several were immediately identified
as catalogued binary stars, including three X-ray binaries
(at J2000 coordinates [+05 29 34 ; -66 53 02], [+15 13 44 ; -59 07 25]
and [+18 16 59 ; -36 16 26])
which are potential candidates for strong gravity wave production.
A binary star (HD 121454) was found at only $46''arc$ from PSR 1553-62.
Several alignments fell close to the $10''arc$ range, although the precise
nature of the corresponding objects still remains to be determined.
These findings indicate that a thorough search of binary stars
along several hundred pulsar lines-of-sight should reveal a
number of interesting alignments in the $10''arc$ range.
Recall that for our purposes only {\em one} good alignment need
be found.

From [1,12] one can show that the amplitude of the modulation $v_{gw}(t)$ is 
\newline\begin{equation}
v_{gw} \approx 1 m/sec \left( {5''arc\over\phi} \right) 
\left( {f_{gw}\over 10^{-3}Hz} \right)^{5/3} 
{\mu\over M_{\odot}} \left( {M\over M_{\odot}} \right)^{2/3}  \ \ ,
\end{equation}\newline
where $\phi$ is the angular separation between pulsar and binary star and
$f_{gw}$ the gravity wave frequency
(twice the orbital frequency). $\mu$ and $M$ are the reduced mass and the 
total mass of the binary, and $M_{\odot}$ is the mass of the Sun.

A desirable candidate (i.e. a combination of pulsar, interstellar medium
and binary star) is also one that involves a stable interstellar medium
(which is in fact the generic case) and a moderate value of $V(t)$
(order of magnitude of $10 km/sec$.)
Under such circumstances, the uncertainty in the measurement of 
${\tilde V}$ can be reduced to $1\%$ in an optimized and 
updated version of the Penticton-Jodrell Bank experiment, 
observing in a bandwidth that is broad enough for several independent
features and possibly
using more than two radio antennae. To achieve this precision without
making each measurement last more than about $10^{3}sec$, one must
observe at the longest possible radio wavelengths, since the scintillation
features are then smaller and their crossing by the Earth proportionally
faster [2].

The $10^{8}sec$ long experiment should then yield a time series
of about $10^{5}$ measurements (normalized 
$[ {\tilde V}(t_{i}) - V(t_{i}) ] = n(t_{i}) + v_{gw}(t_{i}) \ \  v_{gw}<<n$)
that looks like random noise    
with a standard deviation $\sigma \approx 100 m/sec \ $ ($1\%$ of
$10 km/sec$.)
According to eq(2), a gravity wave signal of amplitude 
$v_{gw} > 100/10^{2.5} \approx 0.3 m/sec$ could then be detected
by the virtual trigger mechanism. 
It was confirmed numerically that gravity wave modulations
of $1 m/sec$ (see eq.(3)) or more are detectable (fig.2) and efforts to
approach the $0.3 m/sec$ limit are underway.

The numerical simulation starts with the generation of a random
signal $n(t)$ with correlation time $t_{c}$ and standard deviation
$\sigma$ that is derived from the delta correlated, zero-mean 
Gaussian white noise $G(t)$ through the low-pass filter
\newline\begin{equation}
{dn(t)\over dt} = -{n(t)\over t_{c}} + {\sigma\over {\sqrt t_{c}}} G(t) \  \ .
\end{equation}\newline

A pulse train $p(t)$ is then generated as prescribed in the 
earlier discussion of scheme 1. In evaluating the power spectrum
${\cal P}(\omega)$ of $p(t)$ the difficulty consists in writing
a code that (1) generates a periodigram $P(\omega)$ with a small
enough  variance, and
(2) can accommodate occurrences of missing data due to observational
constraints.

As can be seen in fig.(2), the code was able to detect a simulated
gravity wave signal $v_{gw}$ buried in a noise $n(t)$
that is one hundred times louder ($\sigma = 100 v_{gw}$).
No data folding or periodigram averaging were used. Hence,
there is room for a further increase of the sensitivity ratio
$\sigma/v_{gw}$ beyond the $100$ mark by refining this numerical
phase of the detection.
Current attempts in this direction are underway. The
indication so far is that an additional half an order of magnitude
in sensitivity could be achieved, modulo the dedication of
nontrivial computing resources. 

Finally, it is clear that the virtual trigger mechanism described 
above could be used as a basis for new data processing approaches
in other periodic gravity wave detection projects such as LISA and
partially VIRGO, and more recent proposals based on new applications
of pulsar timing [13] or on astrometry [14,15,16]. It should also
be noted that the sensitivity enhancement implied by eq.(2) can
alternatively be reached by more conventional filter matching
techniques, according to well established data analysis theorems.
Hence, the detectability of the interstellar medium
gravity wave effect is not conditional to the ultimate success
of the new numerical approach.

To conclude, it appears that through the optimization of 
existing techniques in a few different fields of astronomy and 
the dedication of sufficient computing power, one can
design an effective project for a timely detection of
periodic gravity waves. The underlying idea in this approach
is to make new combinations of astronomical and numerical phenomena 
do a substantial part of the detection work.
  
\vspace*{1.cm}
\centerline{\bf Acknowledgements}
\vspace*{0.5cm}

I have greatly benefited from several discussions with W.G Unruh,
including one which put me on the track of adapting stochastic
resonance ideas to gravity wave detection. I am also grateful to
F. Moss for documentation about his pioneering work in nondynamic
stochastic resonance. A.G. Lyne was very helpful in clarifying
some issues in interstellar medium astronomy. T.R. Marsh has helped
repeatedly with information on peculiar binary stars. C.F. Quist
has gracefully communicated his findings prior to publication.
B. Bergersen has provided much encouragement and logistical help. 
This research was supported by the Cosmology Program of the Canadian
Institute for Advanced Research and it made use of data obtained 
through the High Energy Astrophysics Science Archive Research Center 
Online Service, provided by the NASA-Goddard Space Flight Center.
\clearpage
\centerline{\bf References}
\vspace*{1.cm}
[1]  R. Fakir (1994), {\em ``The interstellar medium as a gravity wave detector''} UBC preprint UBCTP-94-011.
\newline
[2] A.G. Lyne and F. Graham-Smith (1990),  
       {\em ``Pulsar astronomy''} (pub.: Cambridge University Press.)
\newline
[3] B.J. Rickett (1993), in {\em ``Wave Propagation in Random Media''}
       (eds.: Tatarski, Ishimaru and Zavorotny; pubs: IOP and SPIE press.)
\newline
[4]  R. Fakir (1993), The Astrophysical Journal, vol.418, 202.
\newline
[5]  R. Fakir (1994), The Astrophysical Journal, vol.426, 74.
\newline
[6] C.F. Quist and R. Fakir (1995), {\em ``Targets for GAP''}, unpublished.
\newline
[7] R. Benzi, S. Sutera and Vulpiani (1981), J. Phys., A14, L453.
\newline
[8] S. Chandrasekhar (1943), Rev. of Mod. Phys., vol.15, 1.
\newline
[9] S.O. Rice (1944), Bell System Technical Journal, vol.23, 282,
and (1945) vol.24, 46.
\newline
[10] R.L. Stratonovich (1963), {\em ``Topics in the theory of random noise''}
(pub.: Gordon and Breach, Science Publishers, Inc., New York.)
\newline
[11] Z. Gingl, L.B. Kiss and F. Moss (1995), 
Europhysics Letters, vol. 29 (3), 191.
\newline
[12]  K.S. Thorne (1987), in {\em ``300 Years of Gravitation''}, S.W.
Hawking and W. Israel, Eds.(Cambridge University Press, Cambridge.)
\newline
[13]  R. Fakir (1994),  {\em Physical Review D, vol.50, 3795}.
\newline
[14] V.V. Makarov (1995), in {\em ``Proceedings of the RGO-ESA Cambridge workshop
on future possibilities for astrometry in space''}, p.117 
(pub.: ESA publications division and ESTEC, Noordwijk, The Netherlands.)
\newline
[15] R. Fakir (1995), {\em ibid} p.113.
\newline
[16] R. Fakir (1996), {\em ``An Astrometric GAP''}, unpublished.

\clearpage
\centerline{\bf Figure captions}
\vspace*{1.cm}
{\em Figure 1.- }
Blowup of the double trigger mechanism of scheme 1
leading from the phase modulated intensity ${\tilde I}(t)$ to
the amplitude modulated intensity-peak duration $\tau(t)$ to
the pulse train $p(t)$, which is to be spectrally analyzed.  
\newline\newline

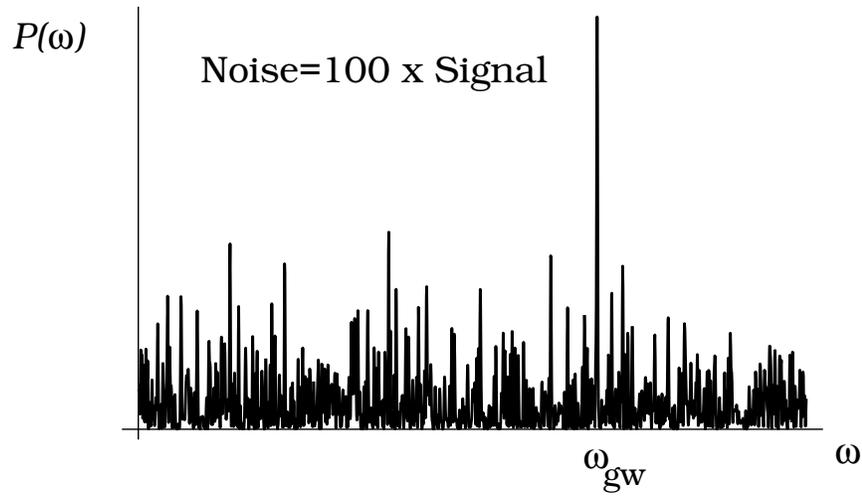
{\em Figure 2.- }
As predicted by eq.(2), the detection (through scheme 2) of a 
gravity wave signal of amplitude $v_{gw}$ and angular frequency
$\omega_{gw}$  buried in random noise of loudness $\sigma$
is successful all the way up to the realistic range $\sigma \sim 100 v_{gw}$.
No period folding or periodigram variance reduction were used.

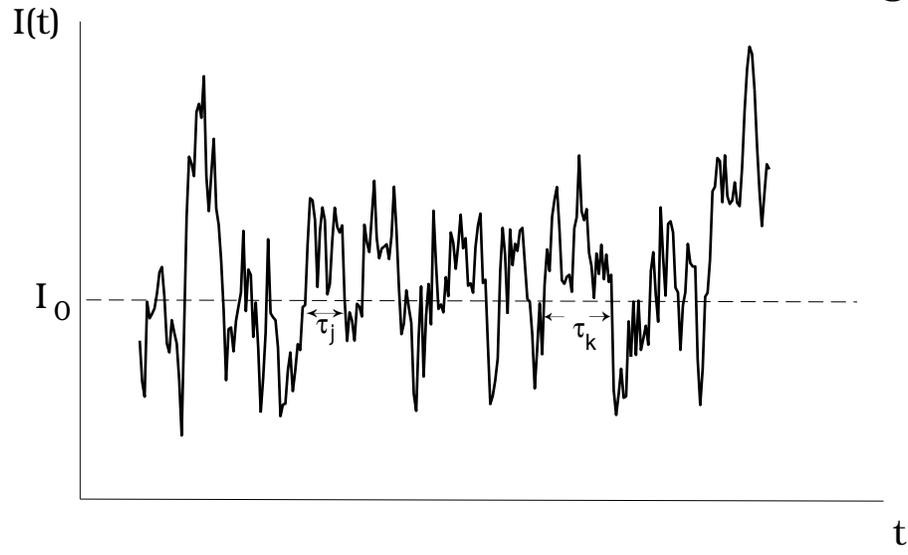
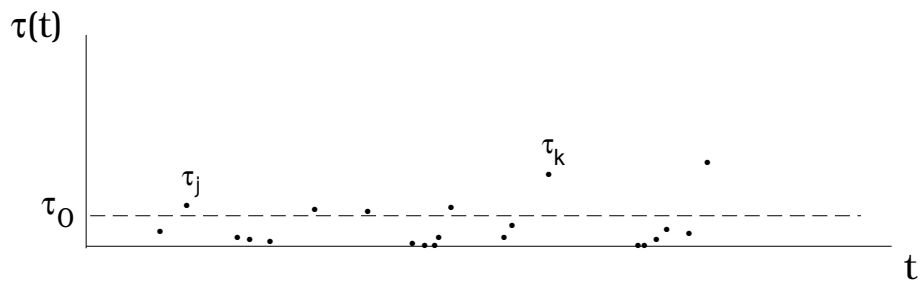
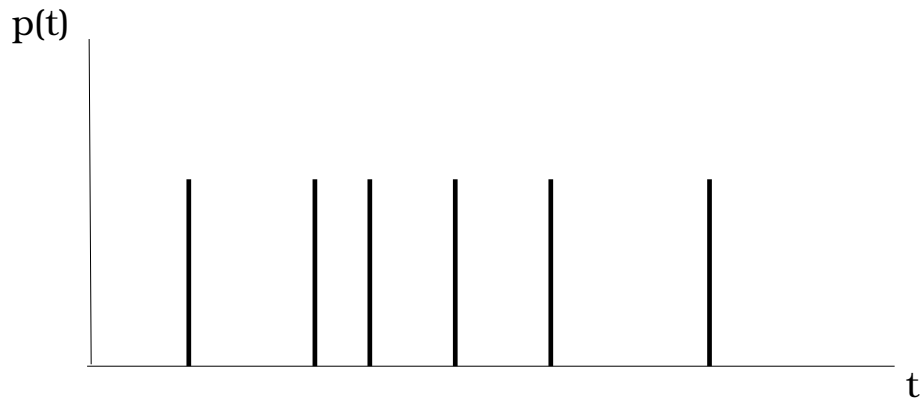
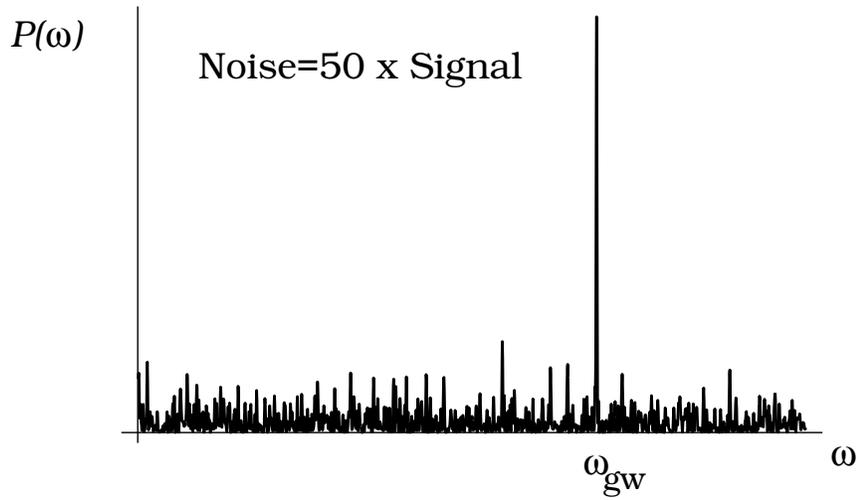
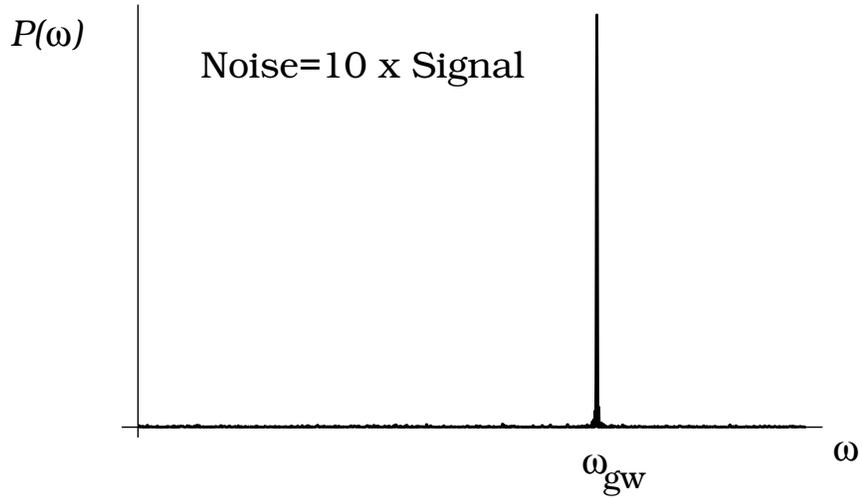
\end{document}